\documentclass{PoS}

\title{Center clusters and their percolation properties in lattice QCD}

\ShortTitle{Coherent center domains in local Polyakov loops}

\author{\speaker{Julia Danzer}, Christof Gattringer\\
        Institute for Physics, Karl-Franzens University Graz, Austria\\
        E-mail: \email{julia.danzer@uni-graz.at} \\ 
\hspace*{10.9mm} \email{christof.gattringer@uni-graz.at}}
                
\author{\vspace{4mm}Szabolcs Borsanyi, Zoltan Fodor\\
        Department of Physics, University of Wuppertal, Germany\\
        E-mail: \email{borsanyi@uni-wuppertal.de} \\
\hspace*{10.9mm} \email{fodor@theorie.physik.uni-wuppertal.de}}

\abstract{\vspace{4mm}
Properties of local Polyakov loops are studied in finite temperature
  lattice QCD and SU(3) lattice gauge theory. We evaluate local Polyakov
  loops,  identify the closest center element for each loop  and investigate
  cluster properties of  these  center phases. For a suitable definition of
  the clusters we find that the deconfinement transition of pure SU(3) gauge
  theory may be characterized by  percolation of the center clusters. For the
  case of full QCD cluster  observables  show a behavior which seems to be
  compatible with a smooth crossover type of  transition.  }

\FullConference{The XXVIII International Symposium on Lattice Field Theory, 
Lattice2010\\
		June 14-19, 2010\\
		Villasimius, Italy}

\begin{document}

\section{Introduction}
With the running and upcoming experiments at LHC, RHIC and GSI the
deconfinement transition of QCD is currently a strong focus of research. We
here report on  our study of percolation aspects in the deconfinement
transition of QCD and SU(3)  gauge theory. Preliminary results were already
presented in \cite{Gattringer,DanzerGatt}. Related studies for SU(2) gauge
theory were reported in \cite{Fortunato1} -- \cite{Alex}.

For analyzing the confinement-deconfinement transition the Polyakov loop  may
be used as an order parameter (for a recent alternative proposal see
\cite{dualcond,Jena}). We here distinguish between local Polyakov  loops
$L(\vec{x})$ located at a spatial point $\vec{x}$ and the spatially  averaged
loop $P = V^{-1}\sum_{\vec{x}}L(\vec{x})$, where $V$ denotes the spatial
volume. The local loop is given by the trace of the product of  temporal gauge
links $U_4(\vec{x},t)$   ($N_t$ is the number of lattice points in time
direction): 
\begin{equation}
L(\vec{x}) \; = \; \mbox{Tr} \; \prod_{t=0}^{N_t-1}\ U_4(\vec{x},t) \;  ,
\label{PloopL}
\end{equation}
i.e., the Polyakov loop is a gauge transporter that propagates a static quark
at position $\vec{x}$ forward in time. The Polyakov loop is related to the 
free energy $F_q$ of a single quark via  $\langle L(\vec{x}) \rangle = 
\langle P \rangle \propto \exp{(-F_q/T)}$, where $T$ is the temperature. 
Below $T_c$ the free energy is infinite, 
$\langle L(\vec{x}) \rangle = \langle P \rangle = 0$, and the quarks are 
confined. Above $T_c$ we have a finite free energy and thus 
$\langle L(\vec{x}) \rangle = \langle P \rangle\ne 0$,
signaling deconfinement. Thus the Polyakov loop acts as an order parameter
for confinement.

The gauge group SU(3) has the three center elements $z=1,e^{\pm i 2\pi/3}$. 
A center transformation with a center element $z$ transforms the temporal 
gauge links at a fixed time slice $t=t_0$: 
$U_4(\vec{x},t_0)\longrightarrow z\ U_4(\vec{x},t_0)$. While
the measure and the gauge action are invariant under the center transformations,
the Polyakov loop transforms non-trivially. A non-vanishing expectation value 
$\langle L(\vec{x})\rangle = \langle P \rangle \neq 0$ thus signals the 
spontaneous breaking of the center symmetry. This
symmetry and its spontaneous breaking are at the core of the Svetitsky-Yaffe
conjecture \cite{SJC} which states that at $T_c$ the system can be described by
a $3D$ effective spin model with an action which is symmetric under the center
group. The spin degrees of freedom are related to the local loops $L(\vec{x})$. 

For such spin systems it is known that aligned spins form local clusters, and that at the critical temperature these clusters start to 
percolate. Various questions arise naturally: 

\begin{itemize}

\item 
Can one identify such characteristic properties of spin systems directly in QCD?
\item
Are these properties important only at $T_c$, or in a 
finite interval of temperatures?  
\item
What happens when one includes fermions which break the center 
symmetry explicitly?
\end{itemize}

\section{Properties of local Polyakov loops}

We study the distribution properties of the local Polyakov loop for quenched as
well as dynamical SU(3) gauge configurations. In the quenched case we use the
L\"uscher-Weisz gauge action with lattice sizes from $20^3 \times 6$ to $40^3
\times 12 $ and temperatures ranging from $T = 0.63\, T_c$ to $1.32\, T_c$
\cite{Gattringer}. For the full theory the configurations have been produced by
the Wuppertal-Budapest group, using a Symanzik improved gauge action and $2+1$
flavors of stout-link improved staggered quarks at physical quark masses
\cite{WB-group1}. The temperatures range from $T = 100$ MeV to $320$
MeV with lattice sizes $18^3 \times 6 , 24^3 \times 6 , 36^3 \times 6 , 24^3
\times 8$ and $32^3 \times 10$, such that finite volume as well as scaling
studies can be performed.

\begin{figure}[t]
\centering
 \hspace*{-2mm}
\includegraphics[height=50mm,clip]{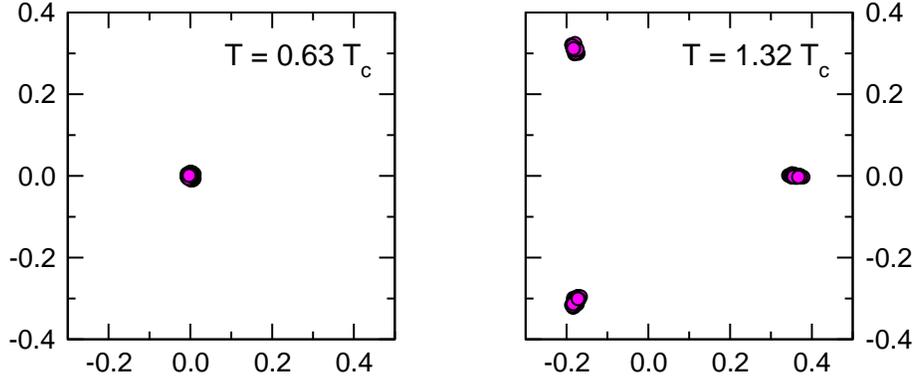}
\caption{\label{Ploop}
Scatter plot of the spatially averaged Polyakov loop $P$ in the complex 
plane for pure gauge theory.}
\end{figure}

Before we come to the analysis of the local loops $L(\vec{x})$ we briefly
summarize the behavior of the spatially averaged loop $P$. This behavior is
illustrated in Fig.~\ref{Ploop}, where we show scatter plots for the values of
the Polyakov loop $P$ in the complex plane. In the lhs.\ plot, which is for a
temperature $T = 0.63 \,T_c$, the values cluster near the origin, i.e.,
$\langle P \rangle = 0$. In the deconfined phase (rhs.\ plot, $T = 1.32
\,T_c$) the values of $P$ are non-vanishing. They scatter at angles $0$ and
$\pm 2\pi/3$, which reflects the underlying center symmetry that is broken
spontaneously above $T_c$. In the infinite volume the system spontaneously
selects one of the three center sectors and only the corresponding ''island''
in the complex plane is populated.  
 
For the analysis of the distribution of the $L(\vec{x})$ we 
write the local loop as
\begin{equation}
L(\vec{x}) \; = \; \rho(\vec{x}) \, e^{i\varphi(\vec{x})} \, .
\end{equation}
We study the histograms $H[\rho(\vec{x})]$ of the distribution of the modulus
$\rho(\vec{x})$, as well as the histograms $H[\varphi(\vec{x})]$ for the phase
$\varphi(\vec{x})$. Analyzing the distribution of the modulus $\rho(\vec{x})$ we
found that both, below and above $T_c$, for quenched as well as for full QCD the
distribution is always the same and closely follows the distribution from Haar
measure \cite{Gattringer,DanzerGatt}.  From that finding we conclude  that the
relevant part of the information must be found in the phase $\varphi(\vec{x})$ 
of the local Polyakov loop. 

\begin{figure}[t]
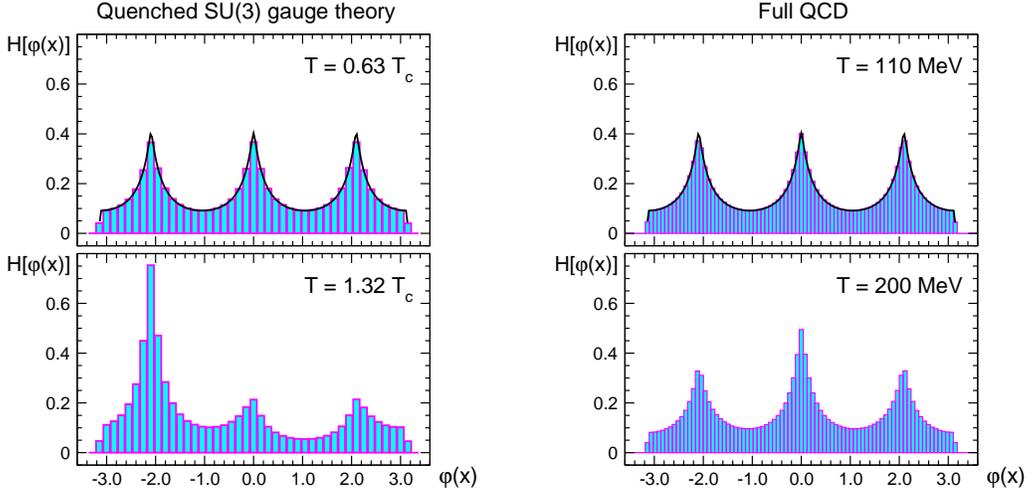

\centering
 \hspace*{-2mm}
\includegraphics[height=65mm,clip]{histo_thetaP_rot_40x6.eps} \hspace{8mm}
\includegraphics[height=65mm,clip]{histo_36x6_phase_dyn.eps}
\caption{\label{phase}
Histograms for the distribution of the local phase $\varphi(\vec{x})$. We
compare quenched (lhs.)  and full QCD (rhs.) at low  and high $T$. The full
curve is the Haar measure distribution 
$H(\varphi) = \int dU \delta(\varphi - \, \mbox{arg}\, U )$.}
\end{figure}

The distribution of $H[\varphi(\vec{x})]$ is shown in Fig~\ref{phase}. In the
quenched case we see that for temperatures below $T_c$ we have peaks in the
distribution of the phase at all three center values. These peaks are equally
populated and average to zero ($1+e^{i2\pi/3}+e^{-i2\pi/3}=0$) when one
calculates the spatially averaged Polyakov loop $P$. 
Above the phase transition the
abundance of sites increases for one of the sectors in a process of spontaneous
symmetry breaking.  For full QCD the situation is similar. For very low
temperatures we have peaks of almost equal height at all three center values.
With increasing temperature the peak of the real sector starts to grow. This is
because the fermion determinant acts like an external field that favors the real
sector. 

The question is now if the phase values at the spatial positions $\vec{x}$ are
distributed independently of each other, or if they form spatial domains as is
known from spin systems. To study this question we assign sector numbers
$n(\vec{x})$ to the sites $\vec{x}$,

\begin{equation}
n(\vec{x}) \; = \; \left\{ \begin{array}{rl}
-1 & \; \mbox{for} \;\;\;\; \varphi(\vec{x}) \, \in \, 
[\,-\pi + \delta \; , \; -\pi/3 - \delta \, ] \; ,\\
0 & \; \mbox{for} \;\;\;\; \varphi(\vec{x}) \, \in \, 
[\,-\pi/3 + \delta \, , \, \pi/3 - \delta \, ] \; ,\\
+1 & \; \mbox{for} \;\;\;\; \varphi(\vec{x}) \, \in \, 
[\,\pi/3 + \delta \, , \, \pi - \delta \,] \; . 
\end{array} \right. 
\end{equation}
If $\varphi(\vec{x})$ is in none of the three intervals no sector number is
assigned to the site $\vec{x}$. $\delta$ is a free real and positive
parameter which allows to cut lattice points  $\vec{x}$ where the corresponding
phase $\varphi(\vec{x})$ is near one of the local minima 
of the distributions  in
Fig.~\ref{phase}. The  remaining lattice points $\vec{x}$ which survive the cut
and are assigned a sector number $n(\vec{x})$ can now be organized in clusters.
We put neighboring lattice  sites $\vec{x}, \vec{y}$ into the same cluster if
$n(\vec{x}) = n(\vec{y})$. 

\begin{figure}
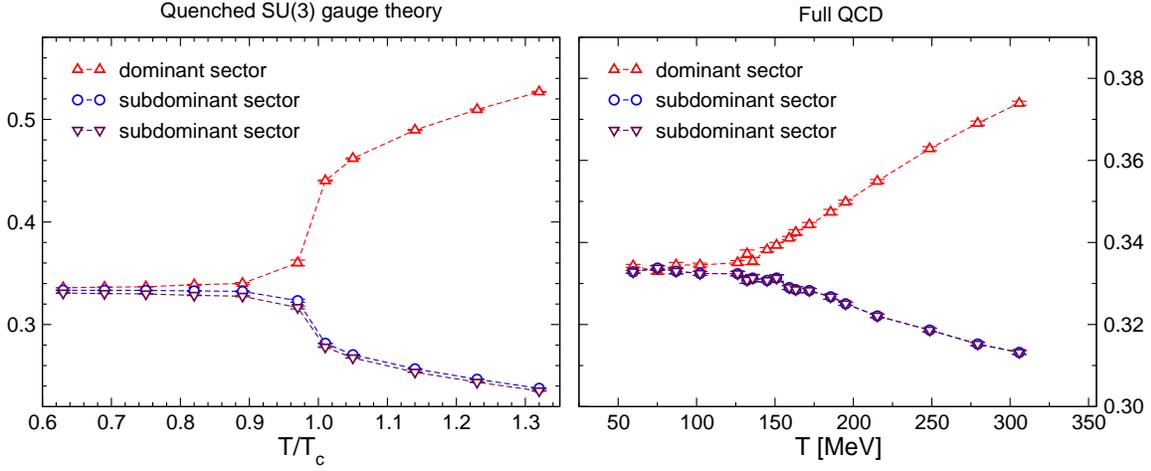

\centering
 \hspace*{-2mm}
\includegraphics[height=62mm,clip]{abundance_vs_ttc_40x6.eps} \hspace{0mm}
\includegraphics[height=62mm,clip]{abundance_vs_ttc_24x8_dyn.eps}
\caption{\label{abundance}
Abundance of lattice sites in the three center sectors as a function of
temperature.}
\end{figure}

As a first test we investigate the number of lattice points 
in the three possible
center sectors as a function of temperature. In Fig.~\ref{abundance} we plot the
occupation of the sectors for the quenched (lhs.) and dynamical data (rhs.) for
$\delta = 0$, i.e., without any cut. We clearly see that in the quenched case
the three sectors are equally populated below $T_c$. At the critical temperature
one of the sectors increases its population while the other two sectors become
depleted.   For full QCD (rhs.) it is the real sector that increases its
population above $T_c$ due to the explicit symmetry breaking from the fermion
determinant. 

\section{Cluster and percolation properties of center domains}

\begin{figure}[b]
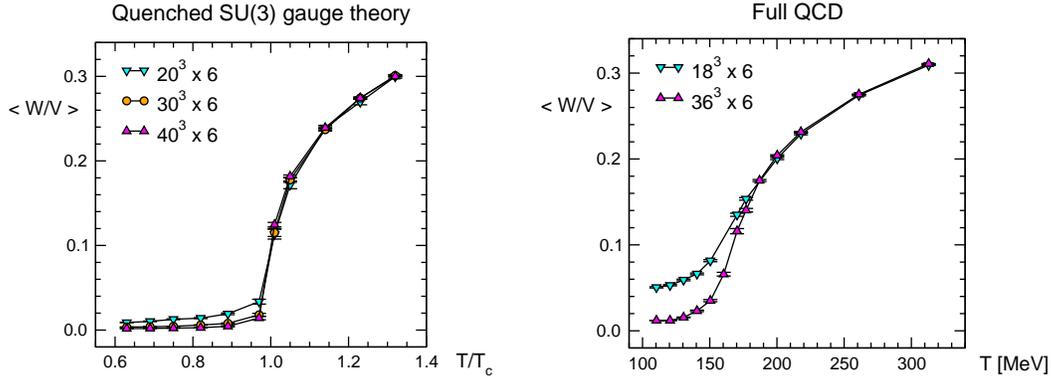

\centering
\includegraphics[height=50mm,clip]{maxcluster_finsize.eps}
\hskip5mm
\includegraphics[height=50mm,clip]{maxcluster_finsize_dyn.eps}
\caption{\label{largest}
Weight $W$ of the largest cluster normalized with the volume $V$ as function of $T$.}
\end{figure}

We now study the dependence of the cluster size on the temperature and
a possible percolation phenomenon of the clusters. For that we plot the
number $W$ of sites in the
largest cluster (''weight of the cluster'') normalized by the spatial volume $V$
as a function of temperature (Fig.~\ref{largest}). For the quenched case 
(lhs. plot) the parameter $\delta$ was set for illustrative purposes such that $39$\% of the sites are cut.
For full QCD a cut of $19$\% was used. The plots show that below $T_c$ 
the clusters are finite with a fixed size, 
which leads to a volume dependence for the ratio $W/V$. 
Above $T_c$ the largest cluster starts to percolate and fills a fixed fraction 
of the volume, as can be seen from the fact that now $\langle W/V \rangle$ is 
independent of $V$. For the quenched case the curves seem to approach a 
non-differentiable limit for $V \rightarrow \infty$, as expected for a
first order transition. In the dynamical case a smooth behavior seems possible.

\begin{figure}[t]
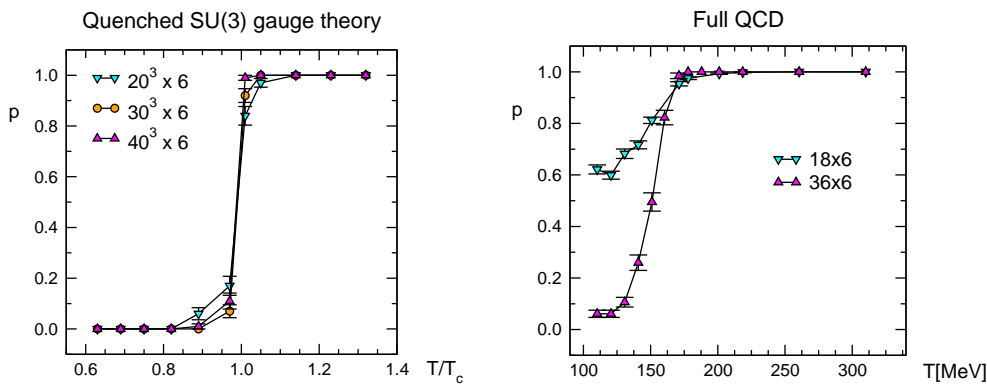

\centering
\includegraphics[height=50mm,clip]{percolation_finsize.eps}
\hskip5mm
\includegraphics[height=50mm,clip]{percolation_finsize_dyn.eps}
\caption{\label{percolation}
Percolation probability $p$ as a function of $T$.}
\end{figure}

In Fig.~\ref{percolation} we plot the probability for finding a percolating 
cluster as a function of temperature. Again the quenched case (lhs.) seems to 
develop a step function, while for the dynamical percolation probability a 
smooth behavior seems feasible. For the latter case a more detailed finite 
volume analysis is necessary for a final conclusion on the
behavior of the clusters near $T_c$.

\section{Summary and outlook}
In the project reported here we explore percolation aspects at the 
deconfinement transition of QCD and SU(3) lattice gauge theory. The starting 
point is an analysis of the phase of the local Polyakov loops which we find
to cluster around the three center values. According to that phase we assign 
the lattice sites to center sectors and study the corresponding center clusters.
For the quenched case we find a sharp onset 
of percolation of the center clusters
at the deconfinement temperature $T_c$. For the case of full QCD the behavior 
seems to be more smooth and could be compatible with the crossover type of
transition expected for full QCD. Additional finite size studies will, however,
be necessary to clearly establish that behavior.    

An important question is whether the clusters and their percolation  properties
can be given a precise meaning in the continuum limit. We have begun to study
this question by comparing results on lattices 
with different lattice  constant $a$.
The cluster definition, i.e., the parameter $\delta$, is set such, that  below
$T_c$ the clusters have a fixed diameter in physical units. 
This  is repeated for
several values of $a$ and the flow of the parameters is monitored. First results
indicate that indeed a continuum limit of the clusters and the percolation
picture seems possible \cite{inprep}. 

\vskip3mm
\noindent
{\bf Acknowledgments:} We thank Christian Lang, Axel Maas and Alexander Schmidt
for interesting discussions. This work was partly supported by the 
DFG SFB TR 55 and the FWF DK 1203.

\end{document}